\documentclass{svmult}

\usepackage{natbib}
\usepackage{makeidx}     % allows index generation
\usepackage{graphicx}    % standard LaTeX graphics tool
\usepackage{multicol}    % used for the two-column index
\bibpunct{(}{)}{;}{a}{}{,}

\newcommand{\ignore}[1]{}
\newcommand{\eq}[1]{\begin{equation} #1 \end{equation}}

\makeindex             % used for the subject index
\begin{document}

\title*{Interaction of Jets with Galactic Winds}
%\titlerunning{Large Scale Simulations of Jets}
\author{Martin Krause%\inst{1}
\and
Max Camenzind}%\inst{1}}
\institute{Landessternwarte K\"onigstuhl
69117 Heidelberg, Germany
\texttt{M.Krause@lsw.uni-heidelberg.de}}
%\and Name and Address of your Institute \texttt{name@email.address}}
\maketitle

We have used the vectorised and parallelised
magnetohydrodynamics code NIRVANA on the NEC SX-5 and the new SX-6 installation in parallel mode
to simulate the interaction of jets with a galactic wind that might be typical for the star-bursting radio-galaxies
of the early universe.

The two simulations, one axisymmetric and one in 3D show that the jet pierces and destroys a thin and dense shell
produced by the pre-installed superwind. 

We suggest that small radio galaxies at high redshift might be absorbed on the blue wing due to the galactic wind shell, 
and possibly on the red wing due to a cooling flow. In larger sources the jet cocoon will fill the wind cavity and accelerate the
shell. The Rayleigh-Taylor instability will then disrupt the shell and disperse dense, possibly star-forming fragments
throughout the region.

\section{Introduction}
\label{intro}
Radio galaxies are associated with an enormous release of energy, exceeding $10^{40}$ Watts \citep{Ghi03,mypap02d}. 
From the power house in the center of certain galaxies, where supermassive black holes of billions of solar masses 
are believed to reside, the power is channeled through narrow jets out to distances of several million light years,
far outside of the host galaxy. They stimulate emission throughout the electromagnetic spectrum that 
can be observed up to the largest distances of the visible universe. Being the likely progenitors of today's
brightest cluster galaxies \citep{Cea01}, radio galaxies at high redshift mark the peaks of highest density in the early universe,
and highlight the formation of galaxy clusters.

At redshifts of $z>0.6$ extended optical continuum and emission line regions, aligned with the radio structures, become prominent companions \citep{MC93}. 
Beyond a redshift of $z \approx 2$, most of the optical emission originates from the Lyman~$\alpha$ transition
of hydrogen. Huge Ly~$\alpha$ halos have been observed in the young universe \citep[e.g.][]{Reulea03}.
Their size often exceeds the radio size. Only for radio extents smaller than 
50~kpc\footnote{kpc = 3260 light years = $3\times10^{19}$ ~m}, the 
Ly~$\alpha$ emission, that has turbulent velocities of typically 1000~km/s, is totally absorbed with 
a velocity width of several~10~km/s \citep{vOea97}. The absorbers are typically blue-shifted \citep{vOea97,dBrea00}.

Optical emission lines are produced by clouds at 
temperatures of $10^4$ to $10^5$~K. These are thought to be embedded in thin and hot plasma in pressure equilibrium.
The emitting gas is either ionised by ultraviolet radiation, e.g. from hot stars or the central quasar, or collisionally excited
due to shocks associated with the jet expansion.
The location of the absorbing screen is still a matter of debate. 
Based on the determination of differing metalicities in the emitting and absorbing gas components,
some authors have suggested an extended low density shell around these systems \citep{Binea00,Jarvea03,WJR03}.
This has the disadvantage that the low velocity width of the absorbers seems to be hard to explain.
Based on hydrodynamic simulation, one of the authors has suggested a different model \citep{mypap02a}.
According to that, the emission line region is surrounded by a high density shell. This shell is formed when the 
gas behind the leading bow shock is dense enough to cool on the propagation time-scale. This part of the gas looses
pressure support and is compressed in a thin and dense shell that would produce the observed absorption.
The latter model may produce too much luminosity in the Ly~$\alpha$ and X-ray channel.
The combination of a jet with a galactic wind is similar to the that model with the exception that lower luminosities
of the absorbing shells are involved.   

The jet--galactic wind interaction scenario 
is detailed in sect.~\ref{detail}, followed by a discussion of the computational aspects in
sect.~\ref{comp}. We present the results of the simulations in sect.~\ref{res}. They are discussed in sect.~\ref{disc}.

\section{Details of the interaction of an extragalactic jet with a galactic wind.}\label{detail}
\begin{figure}
\centering
\begin{minipage}{\textwidth}
\begin{center}
\includegraphics[width=0.47\textwidth]{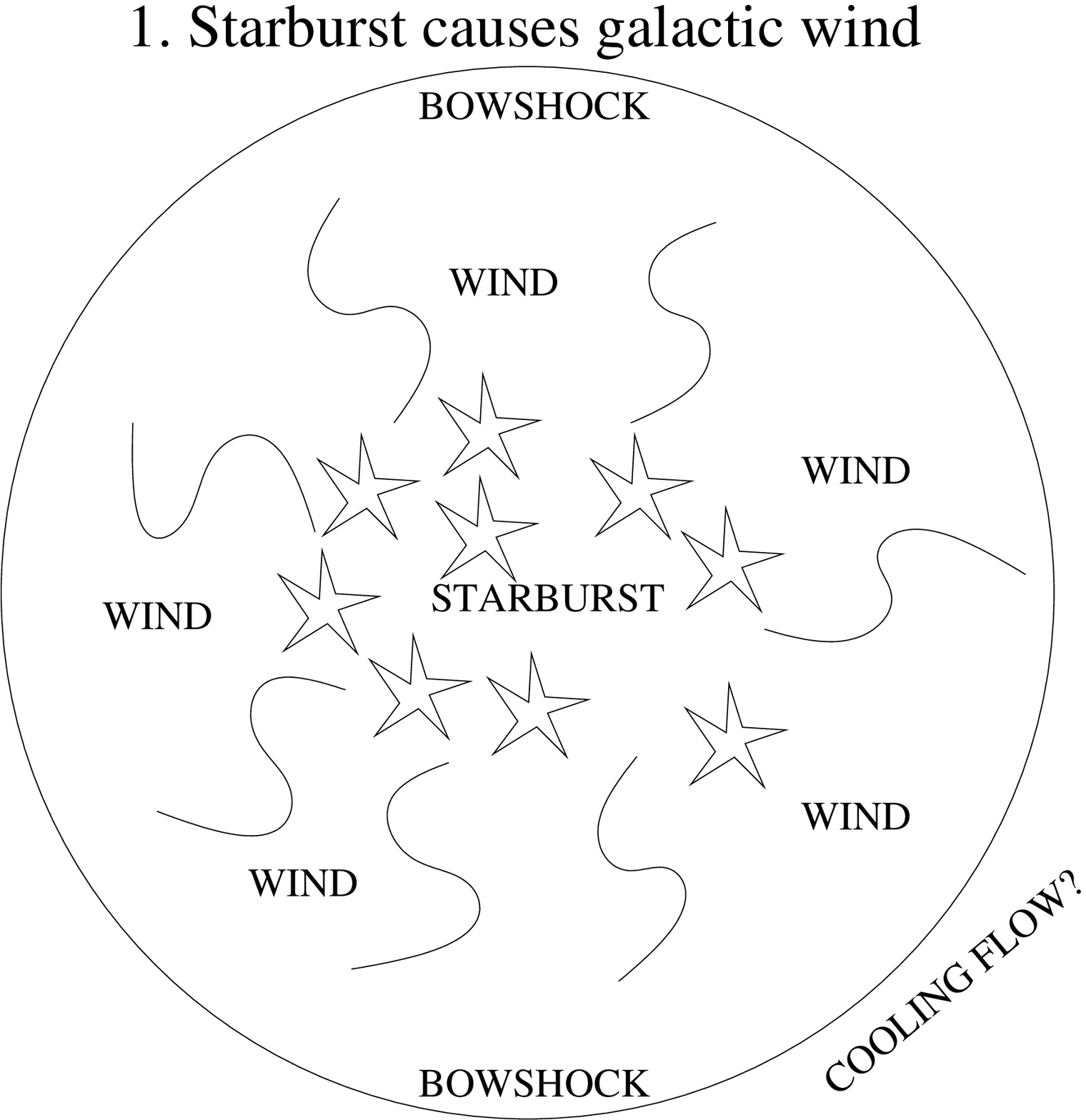}
\includegraphics[width=0.47\textwidth]{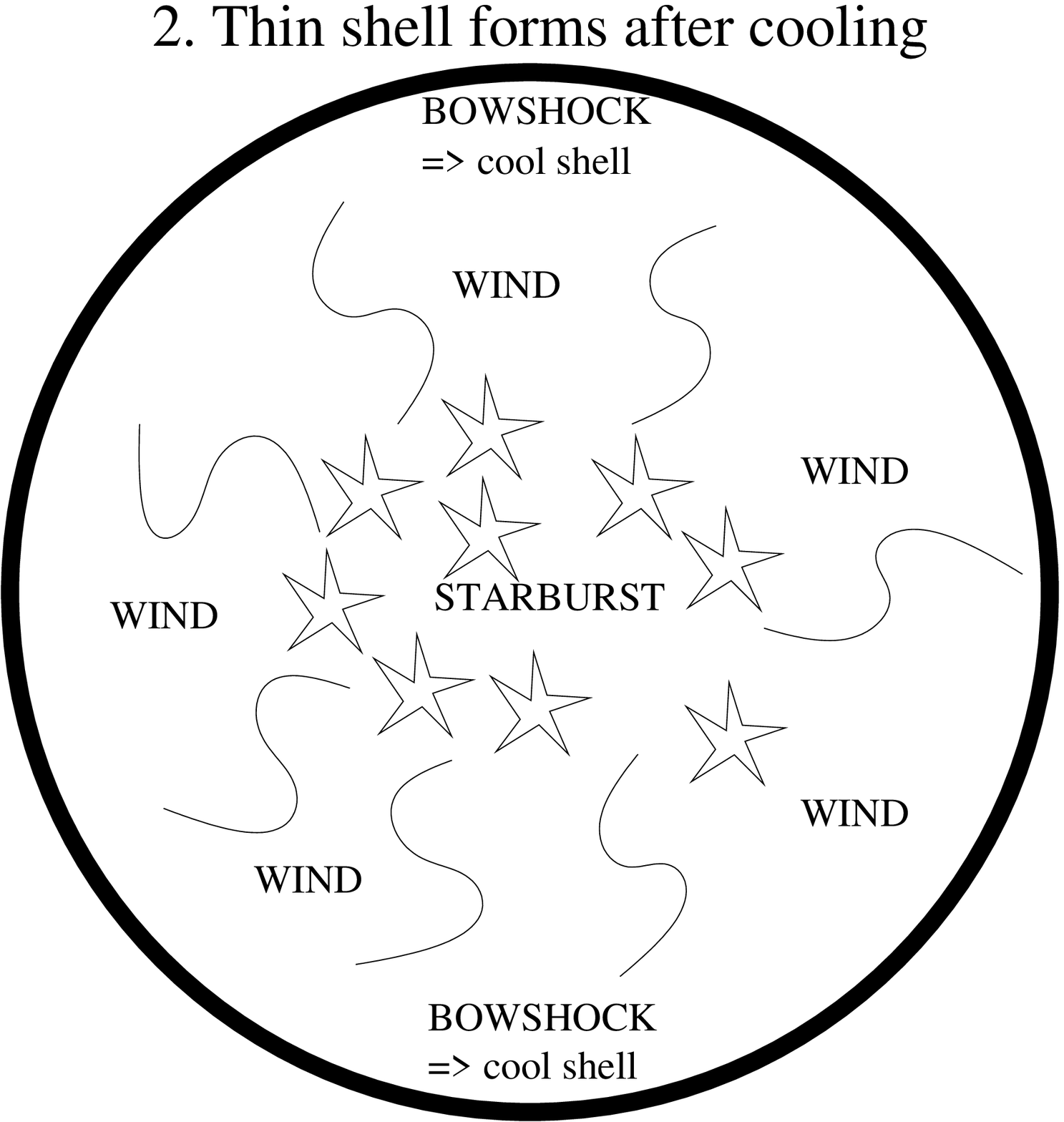}\\
\mbox{}\hfill(a)\hfill\hfill(b)\hfill\mbox{}\\[5mm]
\includegraphics[width=0.38\textwidth]{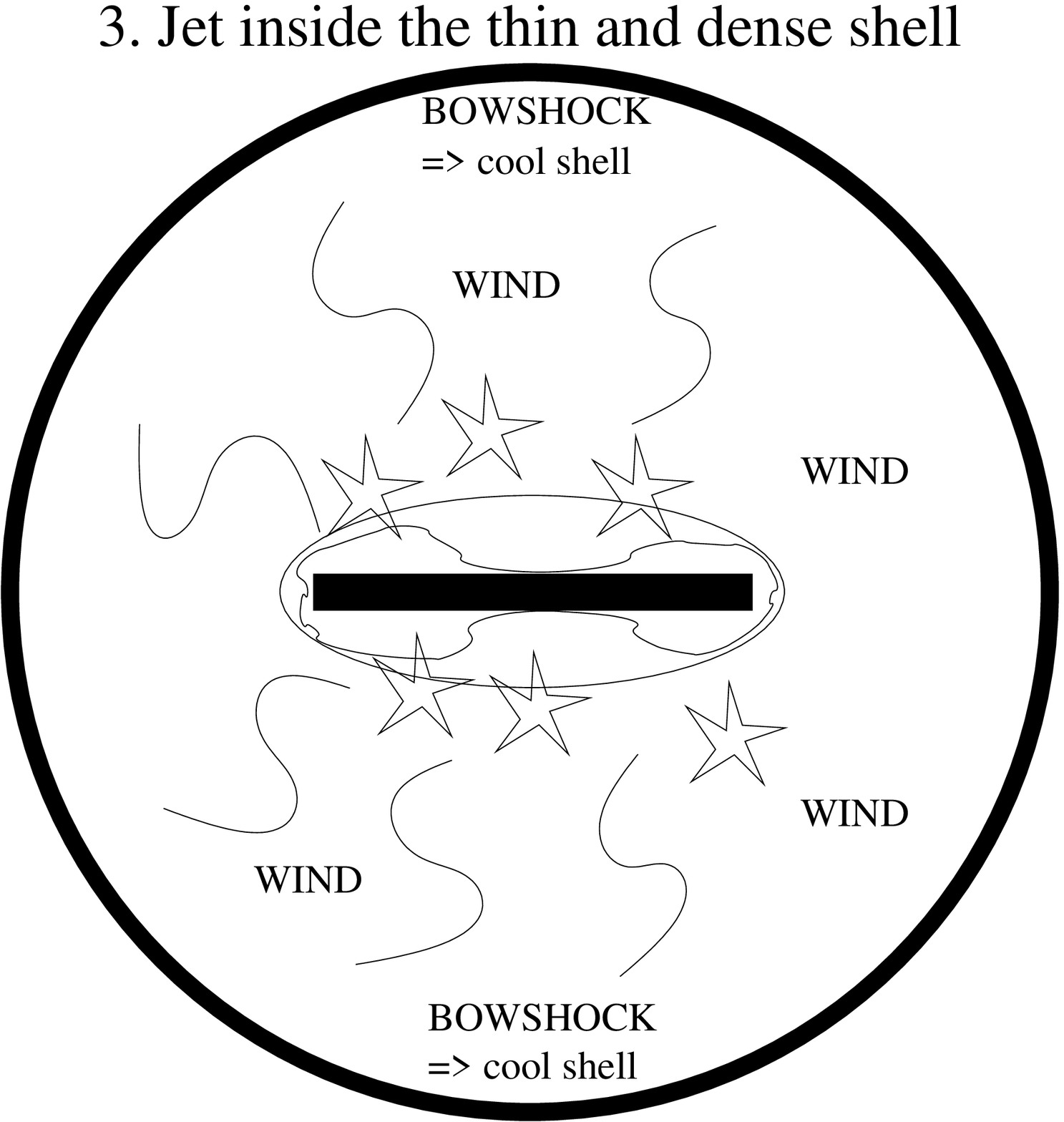}
\includegraphics[width=0.6\textwidth]{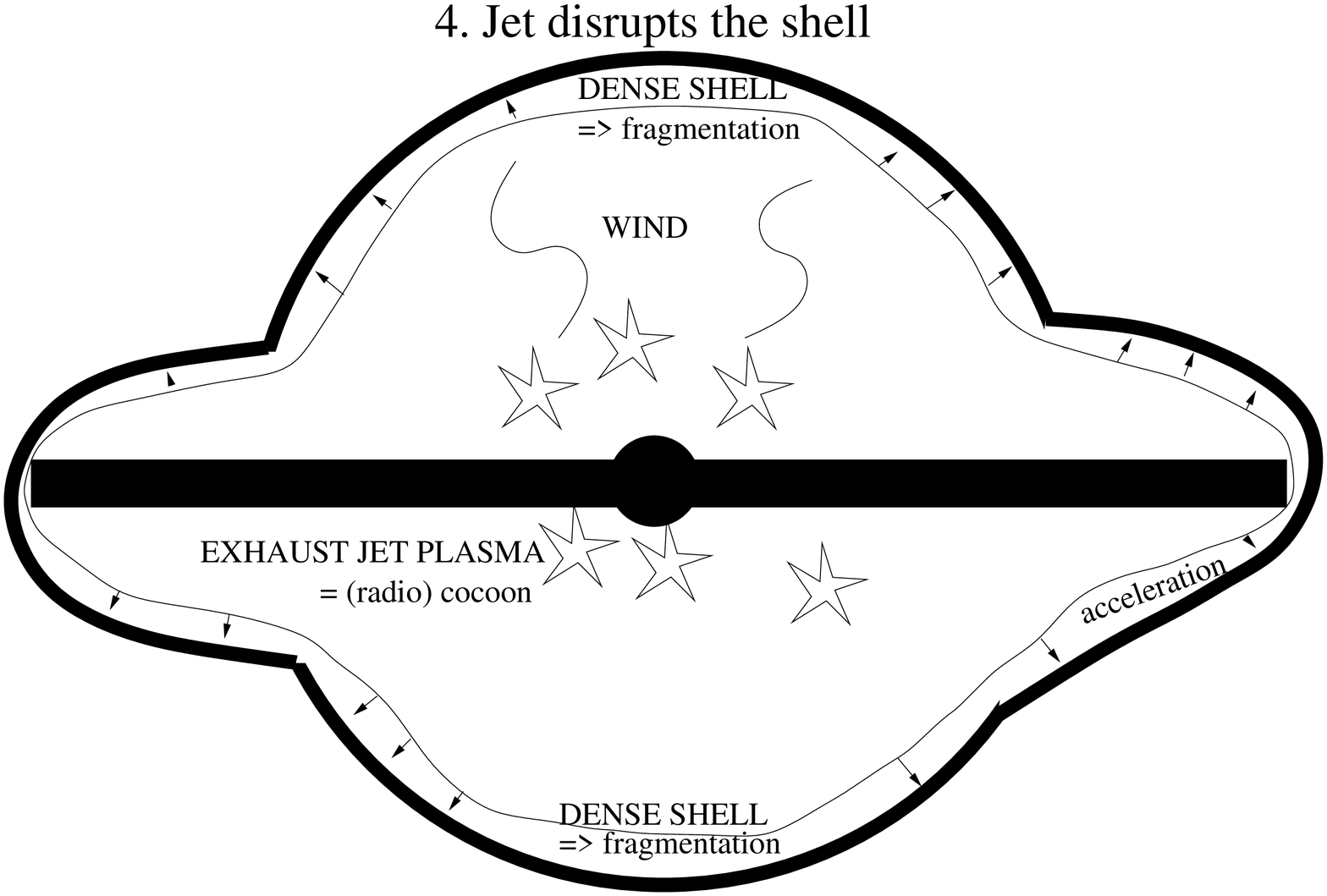}\\
\mbox{}\hfill(c)\hfill\hfill(d)\hfill\mbox{}
\end{center}
\end{minipage}
\caption{Sketch of the interaction of a jet with a galactic wind adopted here.
First, the supernovae associated with a starburst produce a galactic wind (a).
After the cooling time, the gas inside of  the bow shock cools and forms a dense shell (b).
The shell may become Rayleigh-Taylor unstable and loose some material to the inside.
Then, the jet starts (c). The stars still eject gas which forms an emission line region by 
photoionisation or jet induced shocks. The thin and dense shell absorbs at a narrow velocity range.
We suggest the small radio galaxies to be in this phase.
When the jet pierces the shell and the exhaust jet plasma fills the cavity, the shell is accelerated 
because of the high pressure in the interior (d). The shell is then disrupted because of the greatly enhanced
Rayleigh-Taylor instability. The fragments are likely to form stars. 
Radio galaxies bigger than 50~kpc are suggested to be in that phase.
In every phase, a cooling flow may exist that may form an infalling dense shell immediately outside of the wind's bow shock.}
\label{jpw-sketch}       % Give a unique label
\end{figure}

This model (compare Fig.~\ref{jpw-sketch}) 
starts with a starburst. This means that the stellar winds and supernovae inject gas and energy into
the interstellar medium, which leads to an outflow of gas. The outflow is headed by a bow shock (Fig.~\ref{jpw-sketch}a). 
After the cooling time, the material behind this shock will cool and form a dense shell (Fig.~\ref{jpw-sketch}b), 
which may absorb emission
from gas inside of the shell. The shell may be Rayleigh-Taylor unstable and some gas may fall back. Additionally,
ejecta from the stars of the galaxy fill the cavity. When the jet starts (Fig.~\ref{jpw-sketch}c), the gas inside of the cavity may be 
excited to emit strong Ly~$\alpha$ due to photoionisation by the central quasar, which has to be assumed to be active
when it ejects jets, or by shocks associated with the jet expansion. This emission will be absorbed on the blue wing
due to the expanding shell, and also possibly on the red wing due to an infalling cooling flow.
Eventually, the jet plasma will fill all of the cavity (Fig.~\ref{jpw-sketch}d). 
The enhanced pressure will accelerate the bow shock of the wind, 
which merges with the jet's bow shock at that time. This causes the shell to be highly Rayleigh-Taylor unstable,
and the shell will be disrupted quickly. Additionally, the beams pierce it. The shell fragments will fly around 
in a turbulent way and are likely to form stars.

This scenario was examined by hydrodynamic simulations, which is reported on in the following.

\section{Computational aspects}\label{comp}
\subsection{Numerics}
\label{num}
For the computations in this contribution, the magneto-hydrodynamic (MHD) code
{\em Nirvana} was employed \citep{ZY97}. In that version, it solves the MHD
equations in three dimensions (3D) for density $\rho$, velocity $\vec{v}$,
internal energy $e$, and magnetic field $\vec{B}$:
\begin{eqnarray}
\frac{\partial \rho}{\partial t} + \nabla \cdot \left( \rho \vec{v}\right)&
 = & 0 \label{conti}\\
\frac{\partial \rho \vec{v}}{\partial t} + \nabla \cdot\left( \rho \vec{v}
\vec{v} \right) & = &
- \nabla p - \rho \nabla \Phi+ \frac{1}{4 \pi} \left(\vec{B} \cdot \nabla \right) \vec{B}
-\frac{1}{8 \pi} \nabla \vec{B}^2 \label{mom}\\
\frac{\partial e}{\partial t} + \nabla \cdot \left(e \vec{v} \right) & = &
- p \; \nabla \cdot \vec{v} - \rho^2 \Lambda\label{ie}\\
\frac{\partial \vec{B}}{\partial t} & = &
\nabla \times ( \vec{v} \times \vec{B}) \label{ind}\enspace ,
\end{eqnarray}
where $\Phi$ denotes an external gravitational potential and $\Lambda$ is a temperature 
dependent cooling function (zero metals, zero photoionising field) according to \citet{SD93}.
NIRVANA can be characterised by the following properties:
\begin{enumerate}
\item explicit Eulerian time--stepping,
\item operator--splitting  formalism for the advection part of the solver,
\item method of characteristics--constraint--transport algorithm to solve
the induction equation and to compute the Lorentz forces;
\item artificial viscosity has been included to dissipate high--frequency noise
and to allow for shock smearing in case the flow becomes supersonic.
\end{enumerate}

\begin{figure}
\centering
\begin{minipage}{\textwidth}
\begin{center}
\includegraphics[height=\textwidth, angle=-90]{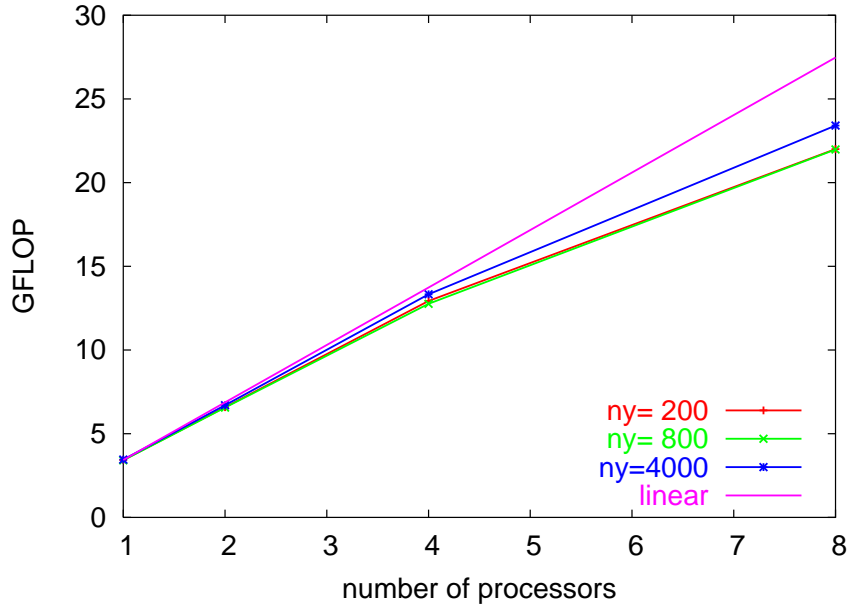}
\end{center}
\end{minipage}
\caption{Scaling of the NIRVANA code on the new SX-6 installation at HLRS}
\label{scaling}       % Give a unique label
\end{figure}

The code was vectorised and parallelised by OpenMP like methods,
and successfully run on the SX-5 \citep{mypap02b}. All the significant loops could be
vectorised. The number crunching part scales without significant performance loss.
This is also true for the MHD part of the solver. 
The code was now modified slightly in order to run on the new NEC SX-6 installation of the HLRS.
A profile output for a test run is shown in table~\ref{tab:1}.
The test run evolved a hydrodynamic problem on a $4096\times4000$ grid for 100 timesteps.
\begin{table}
\centering
\caption{Profile output: SX-6 test run}
\label{tab:1}       % Give a unique label
\begin{tabular}{l@{:}r@{\hspace{0.5cm}}l@{:}r}
\hline\noalign{\smallskip}
  Real Time (sec)    & 192.397907        &     User Time (sec)       &  771.848942         \\
  Sys  Time (sec)    & 2.102633             &    Vector Time (sec)    &  672.636268         \\
  Inst. Count            &  58955512397     &     V. Inst. Count           & 23660611300      \\
  V. Element Count  &   6053551438274 &    FLOP Count              &  2494019257489   \\
  MOPS                   &   7888.650237     &    MFLOPS                  &   3231.227151       \\
  MOPS   (concurrent)   &         57155.158190 &  MFLOPS (concurrent)   &         23411.013720 \\
  A.V. Length           &           255.849325          &
  V. Op. Ratio (\%)      &            99.420335 \\
  Memory Size (MB)      &          2320.000000          &
  Max Concurrent Proc.  &                    8          \\
   Conc. Time($>=1$) (sec)&           106.531878          &
   Conc. Time($>=2$) (sec)&            95.892731\\
   Conc. Time($>=3$) (sec)&            95.830364          &
   Conc. Time($>=4$) (sec)&            95.813671\\
   Conc. Time($>=5$) (sec)&            95.798599          &
   Conc. Time($>=6$) (sec)&            95.782837\\
   Conc. Time($>=7$) (sec)&            95.603244          &
   Conc. Time($>=8$) (sec)&            90.646246\\
  Event Busy Count      &                    0          &
  Event Wait (sec)      &             0.000000\\
  Lock Busy Count       &                17078          &
  Lock Wait (sec)       &             5.299937\\
  Barrier Busy Count    &                    0          &
  Barrier Wait (sec)    &             0.000000\\
  MIPS                  &            76.382190          &
  MIPS (concurrent)     &           553.407238\\
  I-Cache (sec)         &             0.408468          &
  O-Cache (sec)         &             6.734071\\
  Bank (sec)            &            21.394816 \\
\noalign{\smallskip}\hline
\end{tabular}
\end{table}
This output indicates a performance of roughly 23 GFlop for a run with 8 processors
and an acceptable load ballancing.
A single processor run yielded 3.5 GFlops. This is 38\% of the peak performance 
of 9.2~GFlops.  A critical point in achieving this quite good performance was 
a change in the upwind part of the advection solver. 
Now, both directions are computed, and the one needed is selected afterwards.
This saves an if-clause in the vectorised loops.
Although, more floating point operations are performed, the speed increase
is sufficiently high.
The code scales well with the number of processors within one node (Fig.~\ref{scaling}).

We report to computations. An axisymmetric simulation has been performed on one processor of the SX-5,
involving about 200~CPU hours and 200~MB of memory.
Also, a fully 3D computation has been carried out on two processors of the SX-6, involving 100~CPU hours
and 2~GB of memory. The main reason for the short computation time is the limit of 100~CPU hours for the batch queue on
the SX-6 and the fact that the machine was installed only two weeks ago at the time of writing.
We typically need some 1000~CPU hours for a full run. Also this run is currently repeated on the SX-5 and 
evolved for a longer time.

\subsection{Setup}

We start the computations with an isothermal ($10^6$~K) King profile for the initial density $\rho_\mathrm{e}$:  
\begin{equation}\label{kingden}
\rho_\mathrm{e}= 0.3 m_\mathrm{p}
        \left[1+\left(\frac{r}{10 \,\mathrm{kpc}}\right)^2\right]^{-1}\enspace,
\end{equation}
where $r$ is the spherical radius.
The atmosphere is stabilised by the background gravity of the dark matter halo in hydrostatic equilibrium.
We start the galactic wind by injecting mass at a  rate $\dot{M}=10 M_\odot/$yr and thermal energy at a rate 
$L_\mathrm{g}= 10^{51}$~erg/yr, which corresponds to one supernova per year, inside a region of 3~kpc, with an exponential decay. For the 2.5D~run (axisymmetry), the grid has $2047\times1023$ points that correspond to $200\times100$~kpc.
In the 3D case, the grid has $511\times201\times201$ points, corresponding to $200\times80\times80$~kpc.
With a jet radius of one (two) kpc the resolution corresponds to 10 (5) points per beam radius for the
2.5D (3D) case. In both cases, the resolution is enough to resolve the radiative bow shock.

After 80~Myrs, when the dense shell has formed, bipolar jets are injected in the center of the galaxy 
which also corresponds to the center of the grid. The jet density is $10^{-5} m_\mathrm{p}$~cm$^{-3}$.
This is roughly a hundred times less than the gas inside of the wind bubble at the time of the jet start.
The jets are then evolved for 18 (2) Myrs.

\section{Simulation results}\label{res}
\subsection{Axisymmetric simulation}
\begin{figure}
\centering
\begin{minipage}{\textwidth}
\begin{center}
\includegraphics[width=0.35\textwidth]{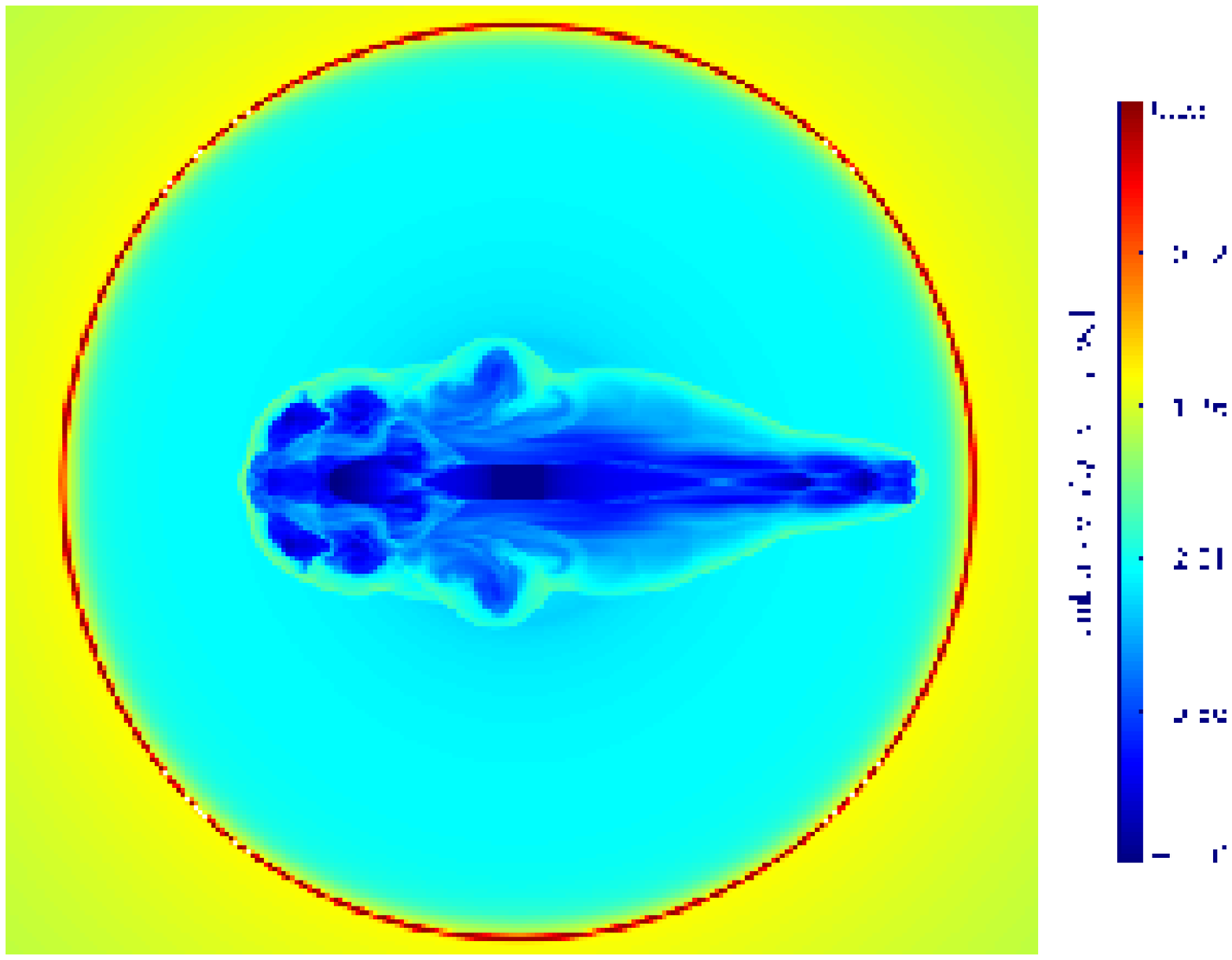}
\includegraphics[width=0.64\textwidth]{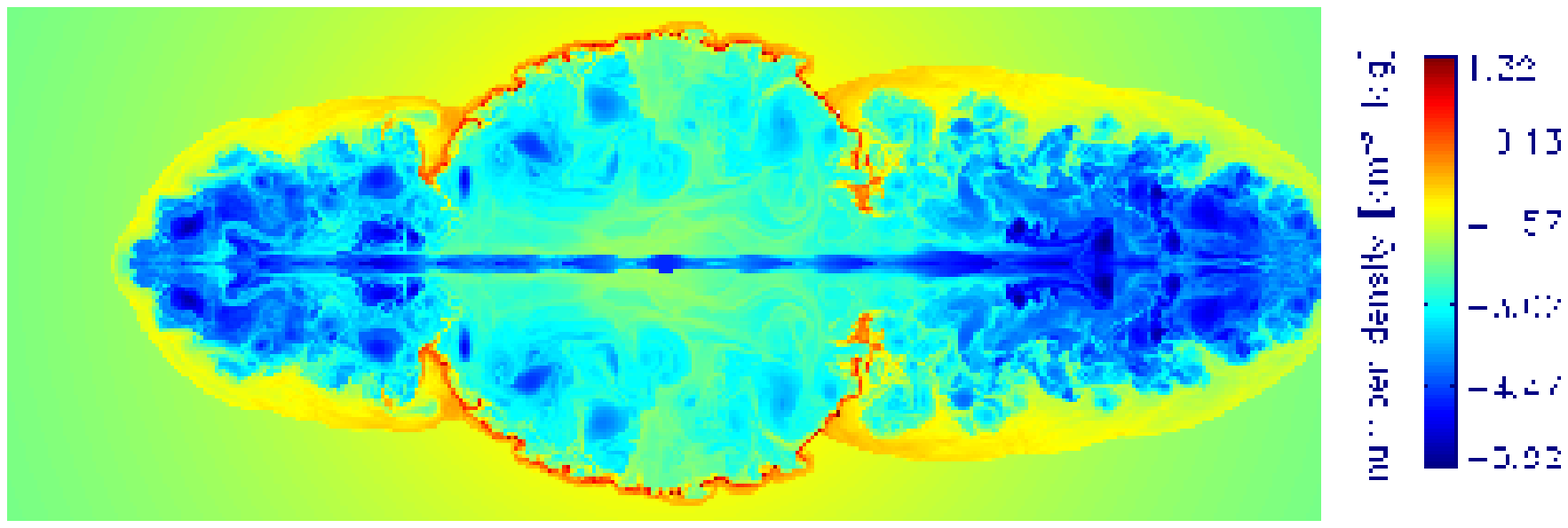}\\
\mbox{}\hfill(a)\hfill\hfill(b)\hfill\mbox{}\\[5mm]
\includegraphics[width=0.35\textwidth]{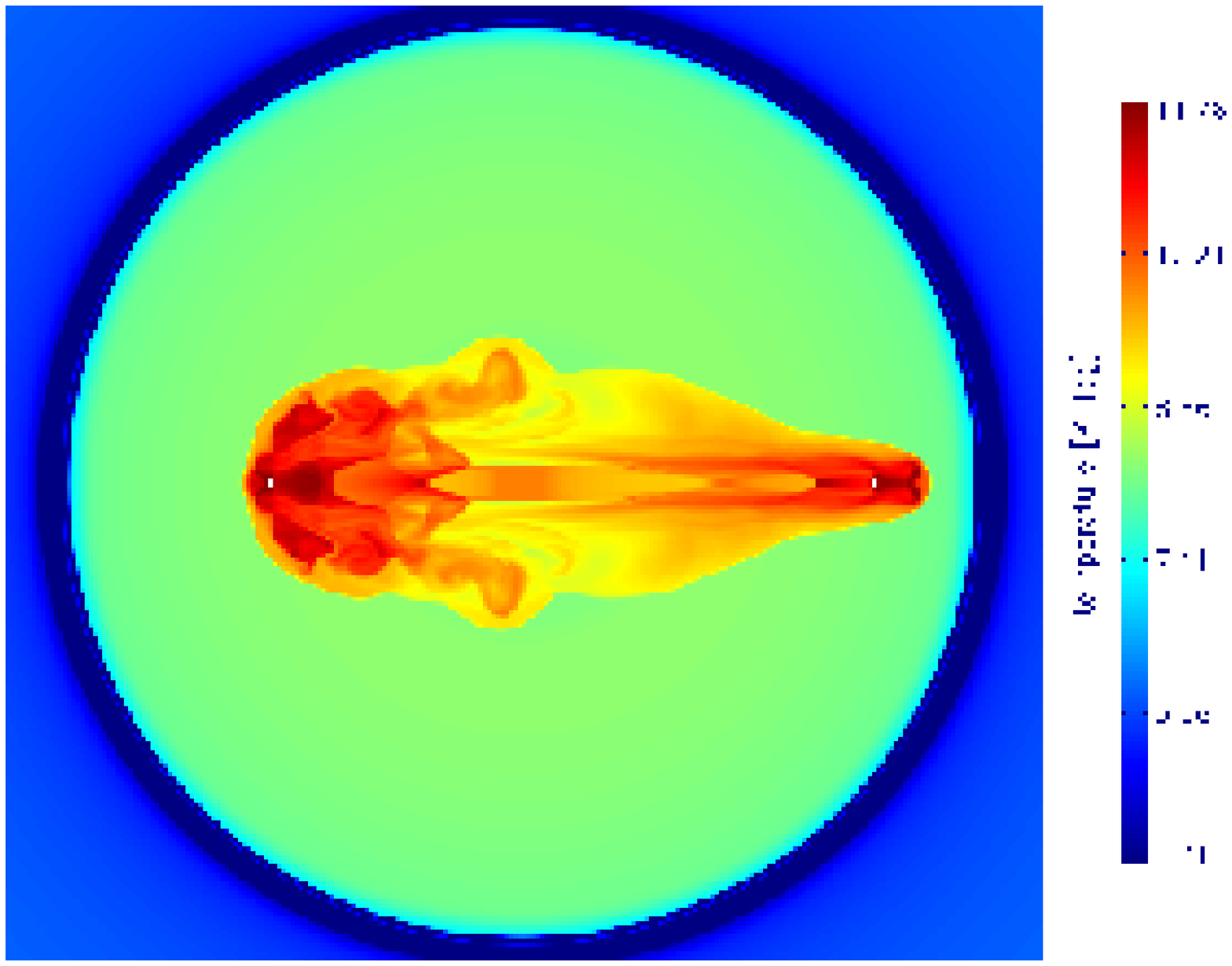}
\includegraphics[width=0.64\textwidth]{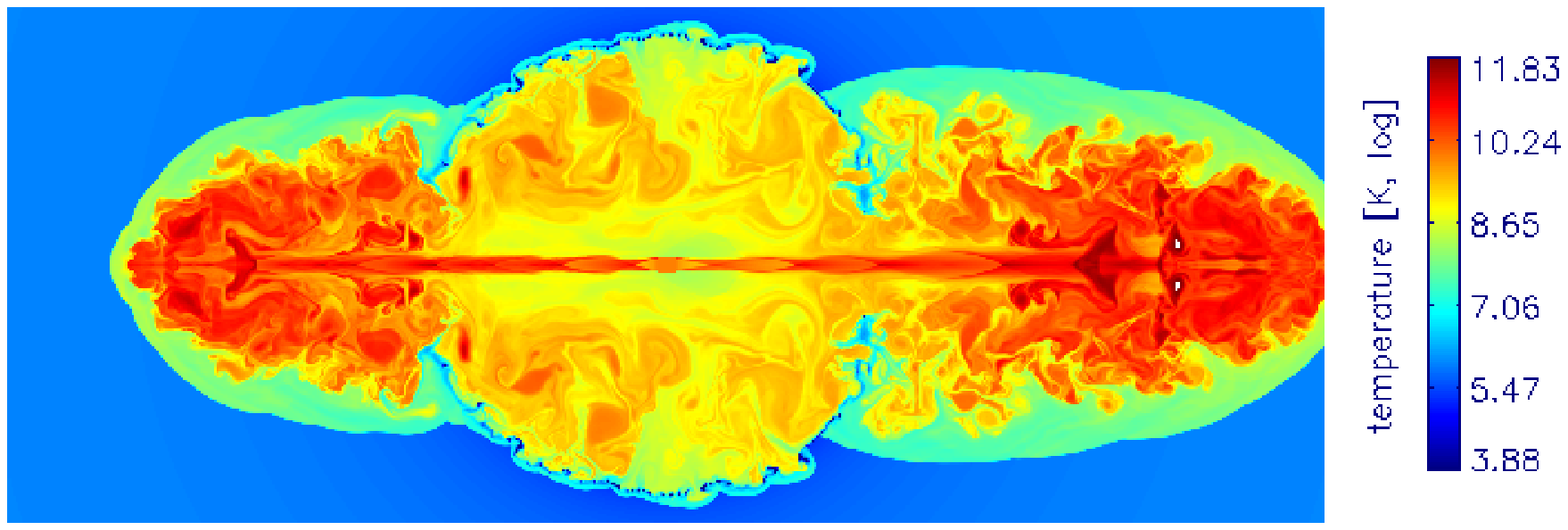}\\
\mbox{}\hfill(c)\hfill\hfill(d)\hfill\mbox{}
\end{center}
\end{minipage}
\caption{Representation of number density (top) and temperature (bottom) 
distribution for the 2.5D run, at 80.5 (left) and 95 Myr (right).}
\label{axi.10}       % Give a unique label
\end{figure}
\begin{figure}
\centering
\begin{minipage}{\textwidth}
\begin{center}
\includegraphics[width=0.99\textwidth]{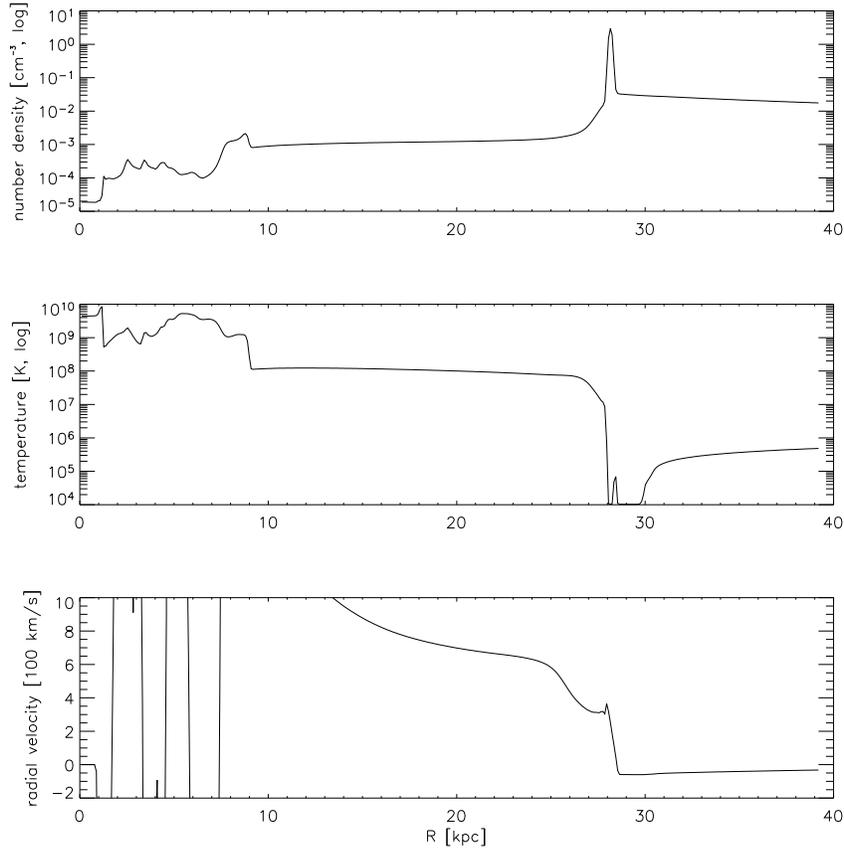}
\end{center}
\end{minipage}
\caption{Slice of density, temperature and sideways velocity for the 2.5D run at 80.5~Myr.
The slice is located vertical to the jet in the center. The bow shock is located at $R=29$~kpc. Inside of it the collapsed 
shell flows outward at $\approx 400$~km/s, outside of it cool gas flows in at $\approx 100$~km/s.}
\label{axi.20}       % Give a unique label
\end{figure}
\begin{figure}
\centering
\begin{minipage}{\textwidth}
\begin{center}
\includegraphics[width=0.99\textwidth]{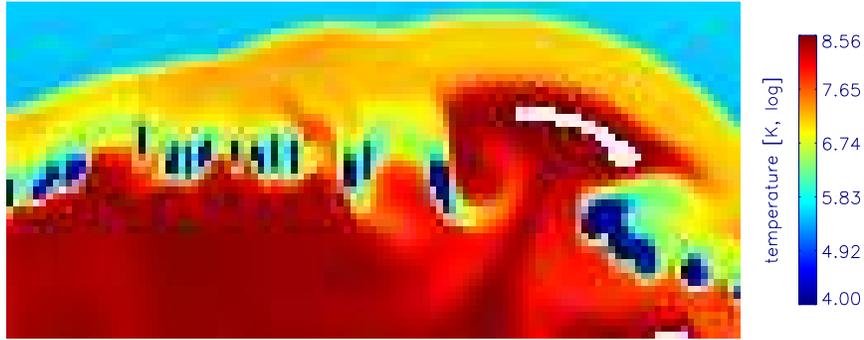}
\end{center}
\end{minipage}
\caption{Magnification of the bow shock region of Fig.~\ref{axi.10}d. The highest temperature has the exhaust
plasma from the jet beam. This plasma (hot, red) flows through the holes between the fragments 
(cold, blue) of the shell and compresses the ambient gas thereby establishing a new,
non-radiative bow shock. The highest temperature is shown in white.}
\label{axi.30}       % Give a unique label
\end{figure}

Density and temperature maps of the 2.5D simulation are shown at two different times in Fig~\ref{axi.10}.
In the earlier plots, the jet is well inside the spherically symmetric galactic wind. This wind has already formed 
a dense shell before the start of the jet. Careful inspection of Fig~\ref{axi.10}c reveals the shell to have a temperature
of $10^4$~K, with a thin enhancement to $\approx 10^5$~K in the middle. This is the position of the bow shock.
The structure can be seen in more detail in Fig~\ref{axi.20}.
The bow shock of the galactic wind expands into a cooling flow. The shock, at 400~km/s, heats the gas to $>10^5$~K.
Inside of the shock, the gas cools quickly down to $10^4$~K, the lower end of the cooling function.
Outside of the shock, the cooling flow has also produced a cool and dense shell, inflowing at $\approx 100$~km/s.
The two jets pierce a considerable hole into these shells. When the jet's bow shock hits the one of the wind,
the shell gets accelerated. This shell can only remain stable as long as its deceleration exceeds the local gravity.
An acceleration greatly enhances the Rayleigh-Taylor instability. Hence, a considerable amount of gas is 
entrained. Fig.~\ref{axi.10}b shows this entrained gas in yellow-green colors, extending in fingers from the shell down
to the jet beam, where it assembles due to gravity. The shell is transformed into clumps that still get denser at the end of the simulation. A magnification of the bow shock region in Fig.~\ref{axi.10}d is shown in Fig.~\ref{axi.30}.
The fragments of the shell can be seen in blue. Between the fragments, the very hot (red) 
jet plasma, that was injected through the 
beams and distributed in the wind cavity, flows past and compresses the ambient gas, which drives a faster, non-radiative
bow shock.

\begin{figure}
\centering
\begin{minipage}{\textwidth}
\begin{center}
\includegraphics[width=1.2\textwidth]{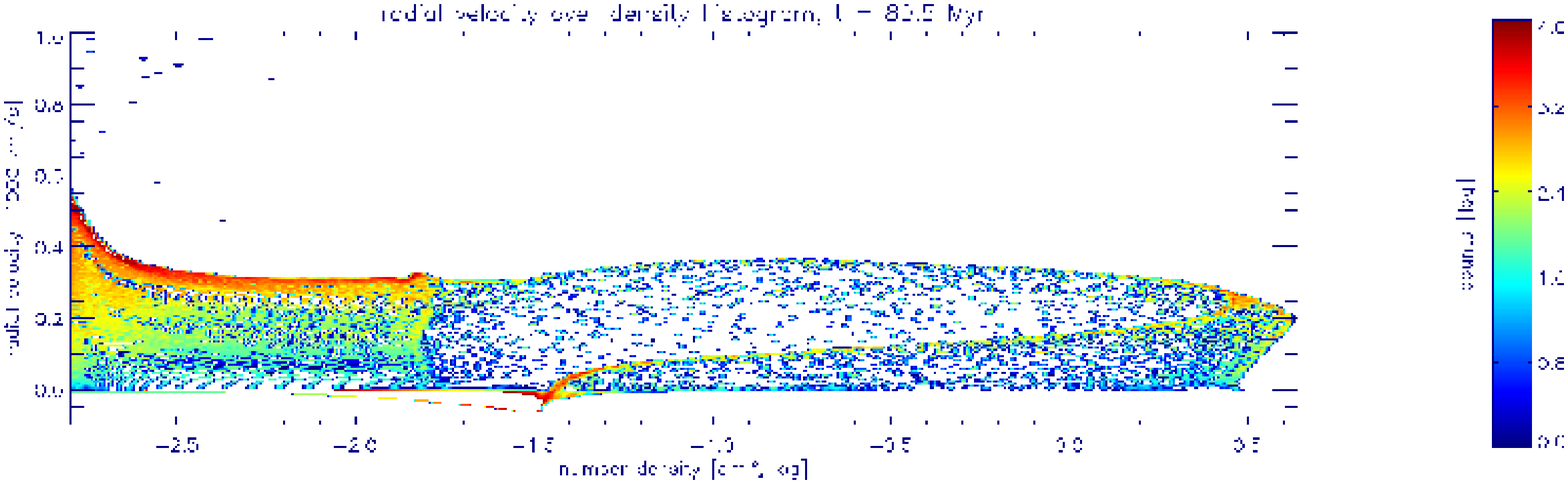}
\includegraphics[width=1.2\textwidth]{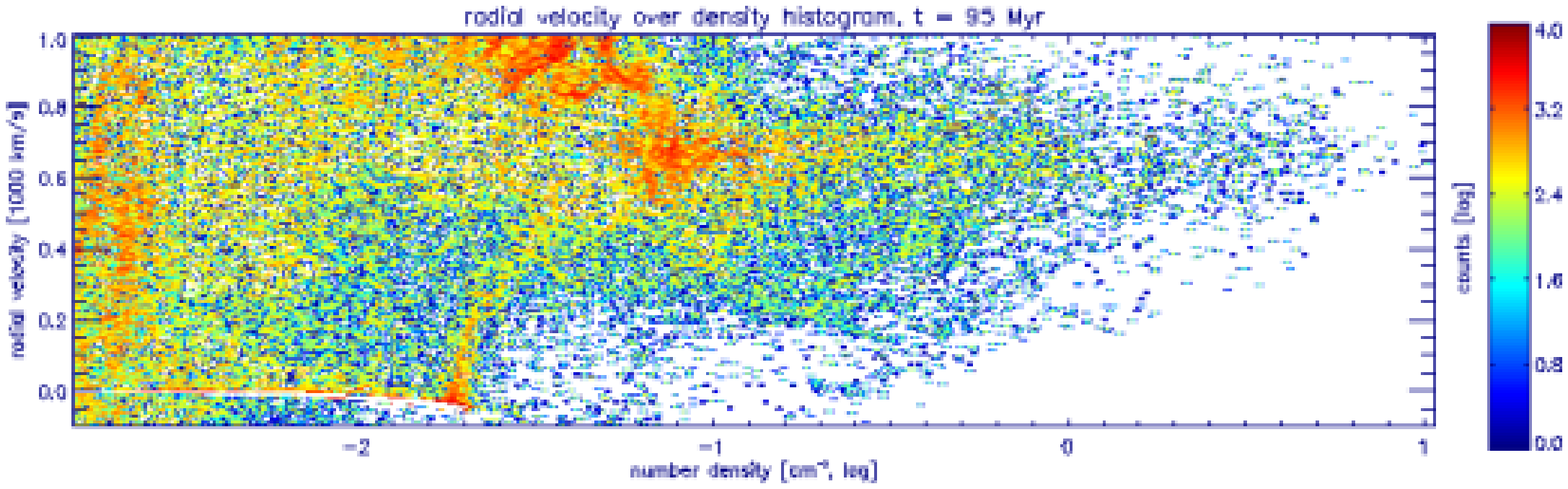}
\end{center}
\end{minipage}
\caption{Frequency of number densities and sideways (radial) velocities at 80.5~Myr and 95~Myr.
The counts are cut at $10^4$.}
\label{axi.40}       % Give a unique label
\end{figure}

The dissolution of the shell can also be seen from the number density versus sideways velocity histograms (Fig~\ref{axi.40}).
At 80.5~Myr, the motion of the dense gas is well ordered, centered at 200~km/s. At 95~Myr, higher densities and higher
velocities are present, but the distribution is much less uniform.

\subsection{Three-dimensional simulation}
\begin{figure}
\centering
\begin{minipage}{\textwidth}
\begin{center}
\includegraphics[width=0.99\textwidth]{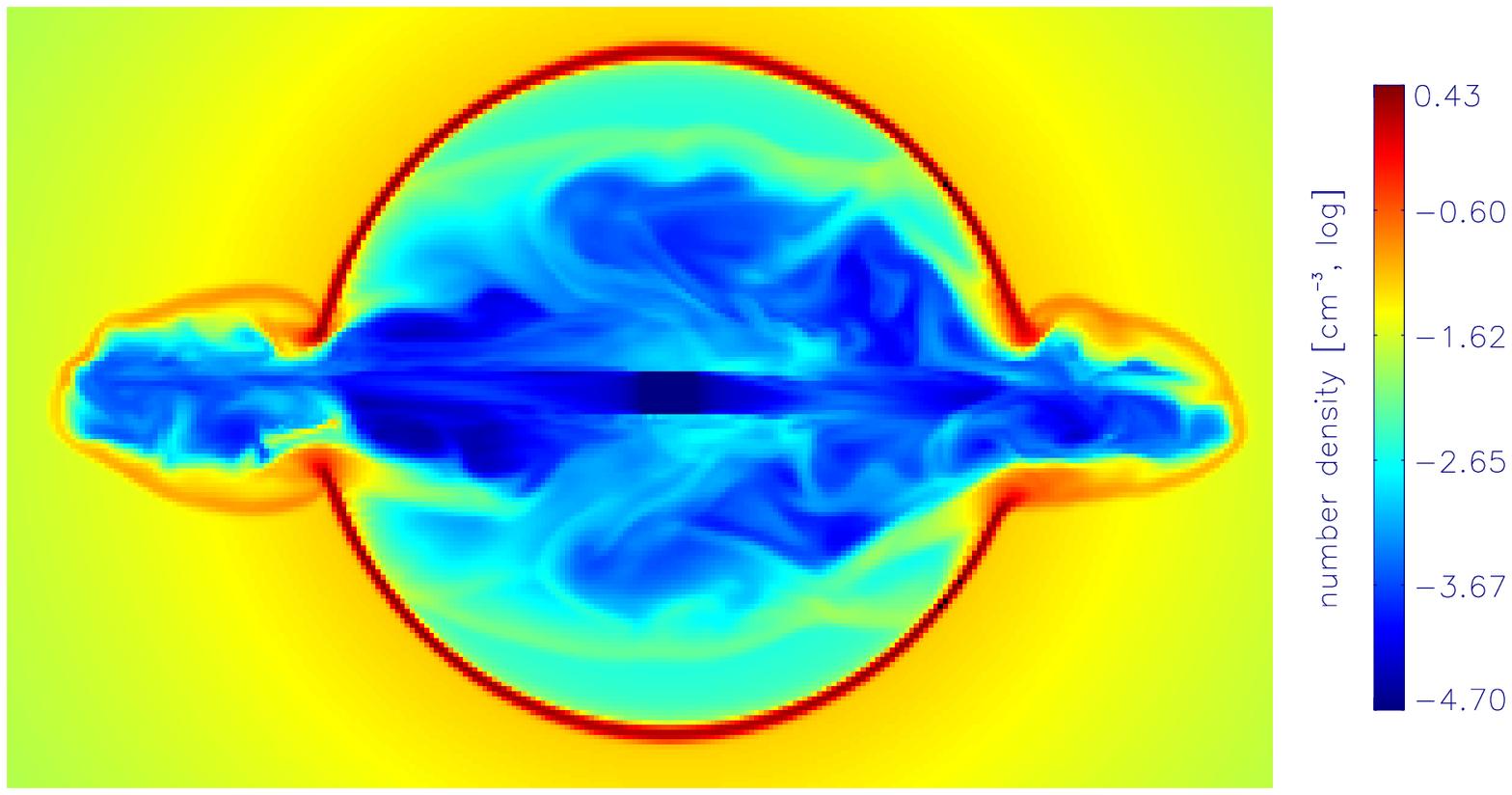}\\
\includegraphics[width=0.99\textwidth]{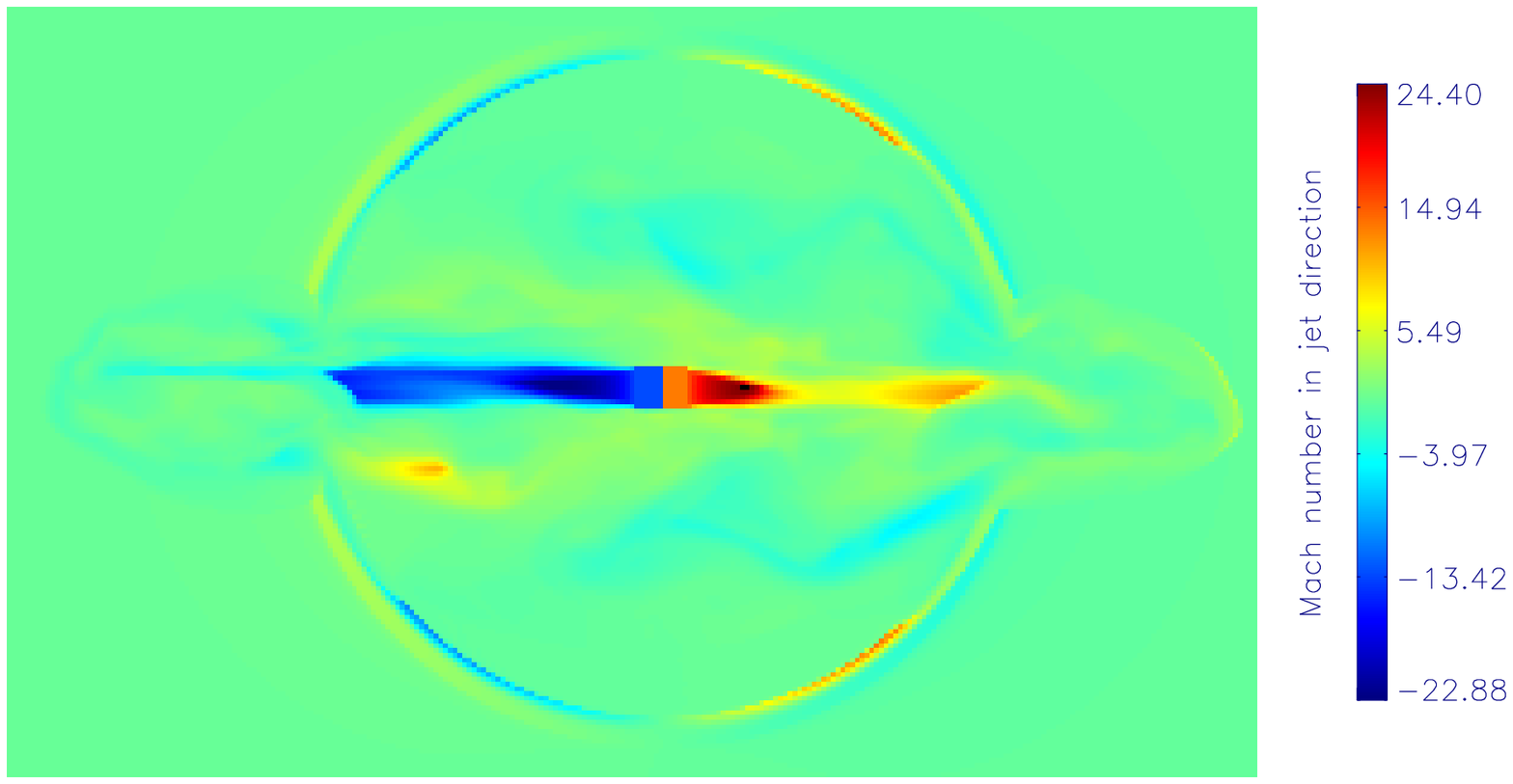}
\end{center}
\end{minipage}
\caption{Density (top) and Mach number in jet direction (bottom) for the 3D~simulation at 82~Myr.}
\label{3d.10}       % Give a unique label
\end{figure}
The jets in the 3D run stay quite symmetric. Only the cocoon shows some three-dimensional structure.
The bow shock of the jet has unfortunately not yet reached the wind's bow shock, and the process of fragmentation can 
therefore not yet be compared to the 2.5D case. An interesting result is that the Mach number drops considerably 
at the location of the shell. 

\section{Discussion}\label{disc}
The simulations show that a jet is able to disrupt a wind blown shell via the Rayleigh-Taylor instability during the simulation time.
This was shown directly in density contour plots, as well as the histograms of velocity versus number density.
The 3D computation has not yet been evolved for long enough in order to compare the fragmentation process.
Another interesting feature to appear in 3D is jet bending. Slight changes of the jet direction might cause the beam to hit 
the shell on one side, which may deflect it. This effect is absent, the beams stay straight.
An emission line region within such a bubble would first be absorbed on the blue wing,
and if a cooling flow is present, possibly also on the red wing.
Observed radio galaxies in the young universe are absorbed preferentially on the blue wing.
This indicates that, if the suggested scenario is correct, the gas around most of these objects is either less 
dense or has a higher temperature, in order not to show a cooling flow. 
However, the properties of the environments of these objects seem to be quite uniform \citep{vOea97}, because
of the strict absence of absorbers for objects larger than 50~kpc, and the presence in nearly all of the objects smaller than that.
It is therefore likely that they are located at the border line. An inflowing cool shell will be created when the cooling time 
for the ambient gas is shorter than the delay between starburst and jet start
($\approx 10^8$ years):
\eq{10^8\,\mathrm{yr}= t_\mathrm{cool,ambient}=12 \,\mathrm{Myr} \,\sqrt{T/10^7\,\mathrm{K}}/(n/\mathrm{cm}^{-3})}
For the gas inside of the wind's bow shock to cool, the shock's Mach number should be close to unity.
Since absorbers are typically blue-shifted by a few hundred~km/s, it follows:
\eq{200\,\mathrm{km/s} \approx c_\mathrm{sound, ambient} = 400\,\mathrm{km/s}\, \sqrt{T/10^7\,\mathrm{K}}}
These two requirements fit together, if the ambient densities are roughly $0.1$~cm$^{-3}$, and the temperature
is typically about $10^6$~K, i.e. the galaxy clusters in the young universe would have had
denser and colder gas than nearby ones.

This comparison, and the simulation results achieved on the NEC~SX-5 and SX~6 supercomputers at the HLRS
suggest that high redshift radio galaxies may indeed be associated with thin shells blown by a galactic wind.

\section*{Acknowledgments}
This work was also supported by the Deutsche Forschungsgemeinschaft 
(Sonderforschungsbereich 439).

% BibTeX users please use
\bibliographystyle{apj}
\bibliography{../texinput/references}

\begin{thebibliography}{14}
\expandafter\ifx\csname natexlab\endcsname\relax\def\natexlab#1{#1}\fi

\bibitem[{{Binette} {et~al.}(2000){Binette}, {Kurk}, {Villar-Mart{\'\i}n}, \&
  {R{\" o}ttgering}}]{Binea00}
{Binette}, L., {Kurk}, J.~D., {Villar-Mart{\'\i}n}, M., \& {R{\" o}ttgering},
  H.~J.~A. 2000, \aap, 356, 23

\bibitem[{{Carilli} {et~al.}(2001){Carilli}, {Miley}, {R{\"o}ttgering}, {Kurk},
  {Pentericci}, {Harris}, {Bertoldi}, {Menten}, \& {van Breugel}}]{Cea01}
{Carilli}, C.~L., {Miley}, G., {R{\"o}ttgering}, H.~J.~A., {Kurk}, J.,
  {Pentericci}, L., {Harris}, D.~E., {Bertoldi}, F., {Menten}, K.~M., \& {van
  Breugel}, W. 2001, in Gas and Galaxy Evolution, ASP Conference Proceedings,
  Vol. 240. Edited by John E. Hibbard, Michael Rupen, and Jacqueline H. van
  Gorkom, San Francisco.

\bibitem[{{De Breuck} {et~al.}(2000){De Breuck}, {R{\"o}ttgering}, {Miley},
  {van Breugel}, \& {Best}}]{dBrea00}
{De Breuck}, C., {R{\"o}ttgering}, H., {Miley}, G., {van Breugel}, W., \&
  {Best}, P. 2000, \aap, 362, 519

\bibitem[{{Ghisellini}(2003)}]{Ghi03}
{Ghisellini}, G. 2003, New Astronomy Review, 47, 411

\bibitem[{{Jarvis} {et~al.}(2003){Jarvis}, {Wilman}, {R{\" o}ttgering}, \&
  {Binette}}]{Jarvea03}
{Jarvis}, M.~J., {Wilman}, R.~J., {R{\" o}ttgering}, H.~J.~A., \& {Binette}, L.
  2003, \mnras, 338, 263

\bibitem[{{Krause}(2002)}]{mypap02a}
{Krause}, M. 2002, \aap, 386, L1

\bibitem[{{Krause} \& {Camenzind}(2002)}]{mypap02b}
{Krause}, M. \& {Camenzind}, M. 2002, in High Performance Computing in Science
  and Engeneering '01, eds.: Krause,~E. and J\"ager,~W., Springer, 329+

\bibitem[{{Krause} \& {Camenzind}(2003)}]{mypap02d}
{Krause}, M. \& {Camenzind}, M. 2003, New Astronomy Review, 47, 573

\bibitem[{{McCarthy}(1993)}]{MC93}
{McCarthy}, P.~J. 1993, \aap Review, 31, 639

\bibitem[{{Reuland} {et~al.}(2003){Reuland}, {van Breugel}, {R{\" o}ttgering},
  {de Vries}, {Stanford}, {Dey}, {Lacy}, {Bland-Hawthorn}, {Dopita}, \&
  {Miley}}]{Reulea03}
{Reuland}, M., {van Breugel}, W., {R{\" o}ttgering}, H., {de Vries}, W.,
  {Stanford}, S.~A., {Dey}, A., {Lacy}, M., {Bland-Hawthorn}, J., {Dopita}, M.,
  \& {Miley}, G. 2003, \apj, 592, 755

\bibitem[{{Sutherland} \& {Dopita}(1993)}]{SD93}
{Sutherland}, R.~S. \& {Dopita}, M.~A. 1993, \apjs, 88, 253

\bibitem[{{van Ojik} {et~al.}(1997){van Ojik}, {R\"ottgering}, {Miley}, \&
  {Hunstead}}]{vOea97}
{van Ojik}, R., {R\"ottgering}, H.~J.~A., {Miley}, G.~K., \& {Hunstead}, R.~W.
  1997, \aap, 317, 358

\bibitem[{{Wilman} {et~al.}(2003){Wilman}, {Jarvis}, {R{\" o}ttgering}, \&
  {Binette}}]{WJR03}
{Wilman}, R.~J., {Jarvis}, M.~J., {R{\" o}ttgering}, H.~J.~A., \& {Binette}, L.
  2003, New Astronomy Review, 47, 279

\bibitem[{{Ziegler} \& {Yorke}(1997)}]{ZY97}
{Ziegler}, U. \& {Yorke}, H.~W. 1997, Computer Physics Communications, 101, 54

\end{thebibliography}
%
% Non-BibTeX users please follow the syntax
% the syntax of "referenc.tex" for your own citations
%\input{referenc}
%%%%%%%%%%%%%%%%%%%%%%%%%%%%%%%%%%%%%%%%%%%%%%%%%%%%%%%%%%%%%%%%%%%%%%

%%%%%%%%%%%%%%%%%%%%%%%%%%%%%%%%%%%%%%%%%%%%%%%%%%%%%%%%%%%%%%%%%%%%%%

\printindex
\end{document}